\documentclass[12pt,letterpaper]{article}

\usepackage{amsmath}
\usepackage{amssymb}
\usepackage{epsfig}
\usepackage{graphicx}
\usepackage{cite}

\newcommand{\capdef}{}
\newcommand{\mycaption}[2][\capdef]{\renewcommand{\capdef}{#2}%
       \caption[#1]{{\footnotesize #2}}}
\makeatletter
\renewcommand{\fnum@table}{\textbf{\tablename~\thetable}}
\renewcommand{\fnum@figure}{\textbf{\figurename~\thefigure}}
\makeatother

\newcounter{myenumi}

\renewcommand{\themyenumi}{\roman{myenumi}}
{\end{list}}

\newlength{\myem}
\newcounter{mysubequation}[equation]

\makeatletter
\renewcommand{\section}{\@startsection{section}{1}{0em}{-\baselineskip}%
{\baselineskip}{\normalfont\large\bfseries}}
\renewcommand{\subsection}%
{\@startsection{subsection}{2}{0em}{-0.7\baselineskip}%
{0.7\baselineskip}{\normalfont\bfseries}}
\makeatother

\newcommand{\sth}{\ensuremath{\sin^2 2\theta_{13}}}
\newcommand{\ldm}{\ensuremath{\Delta m_{31}^2}}
\newcommand{\sdm}{\ensuremath{\Delta m_{21}^2}}
\newcommand{\phase}[2]{\ensuremath{\sin^2 \Delta_{#1}^{#2}}}
\newcommand{\floor}[1]{{\lfloor #1 \rfloor}}

\begin{document}

\begin{titlepage}

\renewcommand{\thefootnote}{\alph{footnote}}

\vspace*{-3.cm}
\begin{flushright}
TUM-HEP-635/06
\end{flushright}

\vspace*{0.5cm}

\renewcommand{\thefootnote}{\fnsymbol{footnote}}
\setcounter{footnote}{-1}

{\begin{center}
{\Large\bf Reactor Neutrino Experiments with a\\[2mm] 
Large Liquid Scintillator Detector}
\end{center}}
\renewcommand{\thefootnote}{\alph{footnote}}

\vspace*{.8cm}
{\begin{center} {\large{\sc
                Joachim~Kopp\footnote[1]{\makebox[1.cm]{Email:}
                jkopp@mpi-hd.mpg.de},~
                Manfred~Lindner\footnote[2]{\makebox[1.cm]{Email:}
                lindner@mpi-hd.mpg.de},~
                Alexander~Merle\footnote[3]{\makebox[1.cm]{Email:}
                amerle@mpi-hd.mpg.de}, \\ 
		and~Mark~Rolinec\footnote[4]{\makebox[1.cm]{Email:}
                rolinec@ph.tum.de}
                              }}
\end{center}}
\vspace*{0cm}
{\it
\begin{center}

\footnotemark[1]${}^,$\footnotemark[2]${}^,$\footnotemark[3]${}^,$\footnotemark[4]
       Physik--Department, Technische Universit\"at M\"unchen, \\
       James--Franck--Strasse, 85748 Garching, Germany

\vspace*{1mm}

\footnotemark[1]${}^,$\footnotemark[2]${}^,$\footnotemark[3]
       Max--Planck--Institut f\"ur Kernphysik, \\
       Postfach 10 39 80, 69029 Heidelberg, Germany

\vspace*{5mm}

\today
\end{center}}

\vspace*{0.3cm}

\begin{abstract}

  We discuss several new ideas for reactor neutrino oscillation 
experiments with a Large Liquid Scintillator Detector. We consider 
two different scenarios for a measurement of the small mixing angle 
$\theta_{13}$ with a mobile $\bar{\nu}_e$ source: a nuclear-powered
ship, such as a submarine or an icebreaker, and a land-based scenario 
with a mobile reactor. The former setup can achieve a sensitivity to
$\sin^2 2\theta_{13} \lesssim 0.004$ at the 90\% confidence level, 
while the latter performs only slightly better than Double Chooz. 
Furthermore, we study the precision that can be achieved for the 
solar parameters, $\sin^2 2\theta_{12}$ and $\Delta m_{21}^2$, with 
a mobile reactor and with a conventional power station. With the 
mobile reactor, a precision slightly better than from current global
fit data is possible, while with a power reactor, the accuracy can 
be reduced to less than 1\%. Such a precision is crucial for testing 
theoretical models, e.g.\ quark-lepton complementarity. 

\end{abstract}

\vspace*{.5cm}

\end{titlepage}

\newpage

\renewcommand{\thefootnote}{\arabic{footnote}}
\setcounter{footnote}{0}

\section{Introduction}

Reactor experiments have always played a crucial role in neutrino physics. The
first experimental evidence for the existence of the neutrino came from a reactor
experiment~\cite{cowanetal1956,cowanetal1956:2} and the most precise measurement 
of the oscillation parameter $\Delta m^2_{21}$ has been performed by the KamLAND 
experiment in Japan~\cite{Eguchi:2002dm}, where the $\bar{\nu}_e$-disappearance 
of neutrinos coming from the surrounding power plants at an average distance of 
approximately~180~km has been studied. KamLAND data has uniquely identified the large 
mixing angle scenario as the correct solution of the solar neutrino problem. 
Furthermore, the non-observation of reactor neutrino disappearance at baselines
$\sim$~1~km in the CHOOZ experiment in France is crucial for the current upper 
bound on the small mixing angle $\sin^22\theta_{13} \lesssim 0.1$~\cite{Apollonio:1999ae}.
The role of future reactor neutrino experiments in measuring or constraining the 
value of $\theta_{13}$ has already been extensively studied in the 
literature~\cite{Minakata:2002jv,Huber:2003pm,Anderson:2004pk,Ardellier:2004ui}. 
Usually it is proposed to construct a relatively small detector close to a large 
nuclear power station. However, it is important to keep in mind that the performance 
of such a measurement besides the systematical uncertainties depends only on the 
total exposure, which is proportional to the product of the thermal power of the 
reactor, the detector mass, and the running time of the experiment. Thus, it is 
also imaginable to take advantage of a bigger multi-purpose detector, such as for 
example the Large Liquid Scintillator Detector LENA~\cite{Undagoitia:1,Oberauer:2005kw}, 
and combine it with a relatively small reactor. In this work, we consider using 
a \emph{mobile} reactor, which will lead to interesting cancellations of systematical 
errors. Mobile reactors are widely used on nuclear-powered ships or submarines, 
but there are also ideas to construct land-based removable power stations that 
may be used to deliver electricity to very remote areas~\cite{hecht}.

Apart from searching for nonzero $\theta_{13}$, an advanced reactor neutrino
experiment can also provide a precision measurement of the solar oscillation 
parameters. It could improve the bounds on $\sdm$, which are currently dominated 
by KamLAND, as well as those on the oscillation amplitude $\sin^22\theta_{12}$,
which are currently dominated by solar data. This improved precision is required, 
for example, to test quark-lepton complementarity~\cite{Minakata:2004xt} to a 
high precision, and to determine the neutrino mass hierarchy in 
$0\nu\beta\beta$-experiments~\cite{Choubey:2005rq,Lindner:2005kr} and through 
the day-night effect in solar neutrino data~\cite{Blennow:2003xw}.
A possible setup for such an advanced reactor neutrino experiment aiming at
a precision measurement of the solar parameters has already been presented in~\cite{sado}.
We will complement the discussion given there by studying a setup involving
a rather small, possibly mobile, nuclear reactor, and a LENA-like detector.
Furthermore we will discuss the possibility of constructing such a large
detector close to a nuclear power station. Since it is usually not desirable to
construct a Large Liquid Scintillator Detector close to such a power plant since
it would then be blind to Geo-neutrinos, supernova relic neutrinos, and other
very faint sources, one would have to choose a reactor which is scheduled for
permanent shutdown after several years of data taking, or a newly built power
plant which only starts operation after the low-background measurements have
been completed.

The outline of the paper is as follows: In Section~2, we will briefly review the
neutrino oscillation framework and the underlying phenomenology of
$\bar{\nu}_e$-disappearance measurements in reactor experiments. In Section~3,
we will discuss the prospects of $\theta_{13}$ measurements with mobile reactors 
for two different scenarios, a nuclear-powered ship and a land-based mobile 
nuclear reactor. We will first describe the expectations from analytical estimates, 
and thereafter present results based on numerical calculations. We carefully include 
systematical uncertainties, which are the main limitation to the achievable 
sensitivity. Next, in Section~4, we will discuss precision measurements of the 
solar parameters $\sin^22\theta_{12}$ and $\Delta m^2_{21}$ with both, a mobile 
reactor and a large nuclear power plant. Again, we will give analytical estimates
as well as numerical results. The latter will include a careful treatment of the
Geo-neutrino background which turns out to have a significant impact on the
optimization of the baseline.

\section{The neutrino oscillation framework}

The relevant oscillation channel for a reactor neutrino experiment is 
$\bar{\nu}_e \rightarrow \bar{\nu}_e$ disappearance. The corresponding exact 
formula for the vacuum survival probability of an electron anti-neutrino 
$\bar{\nu}_e$ with energy $E$ at a baseline $L$ is given by~\cite{Barger:2003qi}:
\begin{multline}
 P(\bar{\nu}_e \rightarrow \bar{\nu}_e)=1-\sth \sin^2 \Delta_{31}
    -(\cos^4\theta_{13}  \sin^2 2\theta_{12}+\sin^2 \theta_{12} \sin^2 2\theta_{13}) \sin^2
     \Delta_{21} + \\
    +\sin^2 \theta_{12} \sth ( \frac{1}{2} \sin 2\Delta_{21} \sin
    2 \Delta_{31}+ 2 \sin^2 \Delta_{31} \sin^2 \Delta_{21} ),
 \label{eq:nuenue_exact}
\end{multline}

where $\Delta_{21}\equiv \sdm L /(4 E)$ and $\Delta_{31}\equiv \ldm L /(4 E)$.
The parameters involved are the ``reactor angle'' $\theta_{13}$, the ``atmospheric 
mass squared difference'' $\ldm$, and the ``solar parameters'' $\theta_{12}$ and 
$\sdm$. The solar parameters have been measured by the SNO and KamLAND 
experiments~\cite{Ahmad:2002jz,Ahmad:2002ka,Ahmed:2003kj,Eguchi:2002dm,Araki:2004mb},
and due to the observed MSW effect~\cite{Wolfenstein:1977ue,Mikheev:1986gs}, the 
sign of $\sdm$ is also known. The atmospheric parameters have been measured by 
Super-Kamiokande~\cite{Fukuda:1998mi,Fukuda:2002pe}, K2K~\cite{Kaneyuki:2005yt,
Ludovici:2006gu}, and MINOS~\cite{Tagg:2006sx,Plunkett:2006sp}, but the sign of 
$\ldm$ has not been determined yet. Considering the small mixing angle $\theta_{13}$, 
there exists only an upper bound which is dominated by the CHOOZ experiment~\cite{Apollonio:1999ae}. 
The best-fit values and allowed intervals for all oscillation parameters from a global 
three-flavor analysis can be found in~\cite{Fogli:2003th,Bahcall:2004ut,Bandyopadhyay:2004da,
Maltoni:2004ei}. 

While our numerical simulations are based on the full three-flavor treatment, 
we will only use zeroth-order approximations for the analytical estimates. 
These approximations are given by
\begin{equation}
P_{\mathrm{sol}}(\bar{\nu}_e \rightarrow \bar{\nu}_e)
   \approx 1-\sin^2 2\theta_{12} \sin^2 \Delta_{21},
 \label{eq:solprob}
\end{equation}

for the measurement of the solar parameters, and by
\begin{equation}
P_{\mathrm{atm}}(\bar{\nu}_e \rightarrow \bar{\nu}_e)
     \approx 1-\sth \sin^2 \Delta_{31}
 \label{eq:atmprob}
\end{equation}

for the (13)-oscillations and are justified by the numerical values of the parameters. 
Since $|\ldm| \gg \sdm$, the (13)-oscillation length is much smaller than the 
(12)-oscillation length, so a typical $\theta_{13}$ experiment located at the 
first atmospheric oscillation maximum is hardly affected by the solar terms. On 
the other hand, for the amplitudes it holds that $\sin^2 2\theta_{12} \gg \sth$, 
so the (13)-oscillation constitutes a negligible perturbation in a measurement 
of $\theta_{12}$ and $\sdm$.

For the oscillation parameters, we assume the following true values in the
numerical simulations: 
\begin{equation}
 \begin{array}{rclrcl}
 \ldm &=& 2.5 \cdot 10^{-3} \textrm{eV}^2, &\hspace{2cm} \sin^2 2\theta_{23} &=& 1,  \\
 \sdm &=& 8.2 \cdot 10^{-5} \textrm{eV}^2, &\hspace{2cm} \sin^2 2\theta_{12} &=&0.83.
 \end{array}
 \label{eq:oscparams}
\end{equation}

\section{$\boldsymbol{\theta_{13}}$ measurement with a mobile reactor}
\label{sec:theta13}

In this section we consider the potential of a mobile electron anti-neutrino source,
combined with a Large Liquid Scintillator Detector, to measure the small mixing 
angle $\theta_{13}$. We assume that the reactor is placed at two different baselines, 
the ``near'' and ``far'' positions, consecutively. The major limitation for such 
an experiment will be the systematical errors. If these are not under control, 
any deficit in the observed neutrino flux could be attributed either to neutrino 
oscillations or to a systematical bias in the initial reactor neutrino flux or 
in the detector properties. We will consider two scenarios with different 
systematical uncertainties:
\begin{itemize}
  \item A nuclear-powered ship, e.g.\ a submarine or an icebreaker (\textit{cf.}~\cite{Detwiler:2002ym}):\\ 
    Since the reactor is not required to be shut down in order to change 
    its position, and since it can be moved between the near and far 
    positions frequently, it is reasonable to assume that the unoscillated 
    neutrino flux and spectrum are the same at both positions. This will 
    turn out to be crucial for the cancellation of the associated 
    systematical errors. Such a nuclear-ship scenario could only be 
    realized, if the detector were located under water, e.g.\ in the 
    Mediterranean Sea. 
  \item A land-based nuclear reactor, such as the SSTAR design~\cite{hecht}:\\
    Here, the reactor needs to be shut down in order to be maneuverable, 
    and such a movement will only be done once, so the neutrino flux
    and spectrum at the near and far positions will be uncorrelated. 
    This problem could be alleviated by using a small dedicated near 
    detector to reduce the uncertainties to those associated with the 
    near detector, which are typically smaller than those associated 
    with the reactor.
\end{itemize}
For both scenarios, we consider a detector with a fiducial mass of 45~kt, which 
corresponds to the proposed LENA detector~\cite{Undagoitia:1,Oberauer:2005kw}.
Note that some of our conclusions can also be applied to reactor neutrino
experiments in which not the source, but the detector is mobile~\cite{goodman}.
In such a scenario, the detector will only be moved once, so it is similar to our 
land-based scenario. However, the detailed impact of the systematical errors
is different: If the reactor is movable, only the neutrino flux and spectrum uncertainties
will be uncorrelated in the two phases of the experiment, while for a movable detector scenario,
also at least some of the detector-related uncertainties should remain uncorrelated.
Apart from this, the feasibility is different since mobile nuclear reactors do
already exist and their development is continuing, while there is no practical
experience on mobile neutrino detectors yet. Note, that only for the nuclear-powered ship
scenario it can be assumed that all neutrino flux and spectrum associated systematical uncertainties are
completely correlated in the two phases of the experiment, since, as mentioned above, a
frequent change between the near and far position is possible. 

To study the sensitivity of reactor neutrino experiments, we use a $\chi^2$ 
analysis, incorporating pull terms for the proper treatment of systematical 
uncertainties. In our numerical calculations we assume the events to follow 
a Poisson distribution, but for illustrative purposes it is sufficient to 
consider the Gaussian approximation, which is very good due to the large event 
rates in a $\theta_{13}$ reactor experiment. For the nuclear-powered ship scenario, 
our $\chi^2$ expression has the form
\begin{multline}
  \chi^2 = \sum_{A = N,F} \sum_i \frac{1}{N_i^A} \big[ T_i^A(a_{\rm norm}, a_{\rm det},
     a_{\mathrm{shape},i}, a_{\rm bkg}, b) - N_i^A \big]^2 +   \\
   + \frac{a_{\rm norm}^2}{\sigma_{\rm norm}^2} + \frac{a_{\rm det}^2}{\sigma_{\rm det}^2}
   + \sum_i \frac{a_{\mathrm{shape},i}^2}{\sigma_{\rm shape}^2}
   + \frac{a_{\rm bkg}^2}{\sigma_{\rm bkg}^2}
   + \frac{b^2}{\sigma_b^2}. \hspace*{2 cm}
   \label{eq:chi2full}
\end{multline}
Here, $N_i^N$ and $N_i^F$ denote the event numbers in the $i$-th bin at the near
and far positions, respectively. These event rates are calculated with the values 
for the oscillation parameters given in Eq.~\eqref{eq:oscparams}. Correspondingly, 
$T_i^A$ are the theoretically predicted event rates for certain fit values of the 
oscillation parameters, and for the systematical biases $a_j$ and $b$. The second 
line of Eq.~\eqref{eq:chi2full} contains the pull terms which represent prior 
knowledge about the systematics parameters. They give a penalty to biases that 
are much larger than the estimated systematical errors. In detail, we introduce 
the following systematical error sources:
\begin{itemize}
  \item The flux normalization uncertainty $\sigma_{\rm norm} = 2.0\%$,
    which is correlated between the near and far positions in the nuclear-powered ship
    scenario, but not in the land-based scenario. It accounts for the limited
    accuracy with which the thermal reactor power and thus the emitted neutrino
    flux can be determined.
  
  \item The detector normalization error $\sigma_{\rm det} = 2.0\%$, which describes
    uncertainties associated with the fiducial detector mass, the cross sections, the
    scintillator properties, and the analysis cuts.

  \item A background flux error $\sigma_{\rm bkg} = 10.0\%$. As backgrounds, we take
    into account Geo-neutrinos, amounting to about 1450~events per
    year~\cite{Hochmuth:2005nh,Geothesis,GeoHP}, and a diffuse reactor background 
    of about 850~events per year, which corresponds to the estimated flux from the 20 
    closest reactors at the proposed LENA site in Pyh\"asalmi (Finland)~\cite{reactors,
    reactors2}. By taking into account all nuclear reactors in the world, the background 
    rate would only increase marginally.
    
  \item The shape uncertainty, $\sigma_{\rm shape} = 2.0\%$, which describes the
    limited knowledge of the reactor neutrino spectrum. The corresponding
    parameter $a_{\mathrm{shape},i}$ is independent for each bin $i$, hence the index
    $i$ is introduced. This parameterization follows the discussion in~\cite{Huber:2003pm}.

  \item An energy calibration error $\sigma_b = 2.0\%$, which is parameterized by $b$.
\end{itemize}
These errors are summarized in Table~\ref{tab:sys}. 
The dependence of the $T_i^A$ on the systematical biases is given by
\begin{table}
  \centering
  \begin{tabular}{lr}
    \hline
    Reactor Neutrino Flux              & 2.0\%  \\ 
    Detector Normalization             & 2.0\%  \\ 
    Detector Energy Calibration        & 2.0\%  \\ 
    Shape Error                        & 2.0\% per bin \\ 
    Normalization of 1\% background    & 10.0\% \\ \hline
  \end{tabular}
  \mycaption{\label{tab:sys} The systematical uncertainties that are assumed 
  in our numerical simulations (based on \cite{Ardellier:2004ui,Huber:2006vr}).}
\end{table}
\begin{equation}
  T_i^A = (1 + a_{\rm norm} + a_{\rm det} + a_{\mathrm{shape},i}) \tilde{S}_i^A(b)
     + (1 + a_{\rm bkg} + a_{\rm det}) \tilde{B}_i^A(b),
\end{equation}
where in turn $\tilde{S}_i^A(b)$ nd $\tilde{B}_i^A(b)$ are the signal and background
rates for the wrong energy binning implied by nonzero $b$. They are obtained
from the correctly binned rates $S_i^A$ and $B_i^A$ according to
\begin{align}
  \tilde{S}_i^A(b) &= (1+b)[ (S_{\floor{\delta}+1} - S_\floor{\delta})(\delta -
  \floor{\delta})
                               + S_\floor{\delta} ],  \label{eq:ecal-S}  \\
  \delta           &= b \cdot (i + t_0 + \tfrac{1}{2}) + i,   \label{eq:ecal-delta}
\end{align}
and a similar expression for $\tilde{B}_i^A$ (see also ref.~\cite{Huber:2003pm}). The quantity
$t_0$ in Eq.~\eqref{eq:ecal-delta} is the energy threshold of the detector,
expressed in units of the bin width. We have used the Gauss bracket $\floor{\cdot}$
to denote the floor function. The expression in square brackets in Eq.~\eqref{eq:ecal-S} 
is essentially a linear interpolation between the events in bin $\floor{\delta}$ and 
those in bin $\floor{\delta} + 1$. If $b$ is not too large, the energy calibration 
never changes by more than the bin width, so that $\floor{\delta} = i$. The factor 
$(1 + b)$ in front accounts for the change of the bin width implied by $b$.

The $\chi^2$ expression for the land-based scenario is similar to 
Eq.~\eqref{eq:chi2full}, but since the reactor flux and spectrum are uncorrelated 
between the near and far positions in this scenario, $a_{\rm norm}$ and 
$a_{\mathrm{shape},i}$ get an additional index $A = N,F$.

\subsection{Analytical estimates}

We will now show that, for the nuclear-powered ship scenario, the most important systematical 
errors can be eliminated in the actual measurement of $\theta_{13}$ if the reactor 
is positioned at two different baselines consecutively. To quantify the performance 
of the experiment, we consider the sensitivity to \sth, which is defined as the 
limit that can be set on \sth\ assuming that the true value is~0. This quantity 
can be calculated by comparing the simulated event rates for non-zero test values 
of \sth\ with those for $\sth=0$ in a $\chi^2$~analysis.

For an analytical estimate, we consider a simplified version of Eq.~\eqref{eq:chi2full},
keeping only the reactor flux error $\sigma_{\rm norm}$ and the detector normalization 
error $\sigma_{\rm det}$, but neglecting spectral uncertainties, energy calibration 
errors, and backgrounds:
\begin{equation}
  \chi^2 = \sum_{A=N,F} \sum_{i=1}^n \frac{1}{N_i^A} \left[
          N_i^A (1 + a_{\rm norm} + a_{\rm det})
                              (1 - \sth \phase{i}{A}) - N_i^A \right]^2
       + \frac{a_{\rm norm}^2}{\sigma_{\rm norm}^2}
       + \frac{a_{\rm det}^2}{\sigma_{\rm det}^2}.
  \label{eq:chi2norm}
\end{equation}
Here, the index $i$ runs over all $n$ energy bins, while $A$ takes the values $N$ 
and $F$ for the near and far baselines, respectively. $N_i^A$ is the total event 
rate in the $i$-th bin at baseline $L^A$ without oscillations, and $\Delta_i^A = 
\ldm L^A / 4 E_i$ is the oscillation phase for baseline $L^A$ and energy $E_i$.

Since the systematical biases $a_{\rm norm}$ and $a_{\rm det}$ as well as the 
oscillation amplitude $\sth$ are small, Eq. (\ref{eq:chi2norm}) can be 
approximated by
\begin{equation}
  \chi^2 = \sum_{A=N,F} \sum_{i=1}^n N_i^A \left(
          a_{\rm norm} + a_{\rm det} - \sth\phase{i}{A} \right)^2
     + \frac{a_{\rm norm}^2}{\sigma_{\rm norm}^2}
     + \frac{a_{\rm det}^2}{\sigma_{\rm det}^2}.
\end{equation}
$a_{\rm norm}$ and $a_{\rm det}$ are fitted by minimizing $\chi^2$, therefore we calculate
\begin{align}
  \frac{\partial \chi^2}{\partial a_{\rm norm}} &=
           \sum_{A=N,F} \sum_{i=1}^n 2 N_i^A \left( a_{\rm norm} + a_{\rm det}
           - \sth \phase{i}{A} \right) + \frac{2 a_{\rm norm}}{\sigma_{\rm norm}^2},
  \label{eq:aNormDeriv} \\
  \frac{\partial \chi^2}{\partial a_{\rm det}} &=
           \sum_{A=N,F} \sum_{i=1}^n 2 N_i^A \left( a_{\rm norm} + a_{\rm det}
           - \sth \phase{i}{A} \right) + \frac{2 a_{\rm det}}{\sigma_{\rm det}^2}.
  \label{eq:aDetDeriv}
\end{align}
By requiring the expressions (\ref{eq:aNormDeriv}) and (\ref{eq:aDetDeriv}) to be 
zero and taking their difference, we obtain
\begin{equation}
  a_{\rm norm} = \frac{\sigma_{\rm norm}^2}{\sigma_{\rm det}^2} a_{\rm det}.
\end{equation}
Substituting this back into Eq.  (\ref{eq:aNormDeriv}) or (\ref{eq:aDetDeriv}), and
assuming the $N_i^A$ to be very large, the systematical biases can be expressed as
\begin{align}
  a_{\rm det}  &= \frac{\sth}{1 + \sigma_{\rm norm}^2 / \sigma_{\rm det}^2} \cdot
           \frac{\sum_{A=N,F} \sum_{i=1}^n N_i^A \phase{i}{A}}
                {\sum_{A=N,F} \sum_{i=1}^n N_i^A},  \\
  a_{\rm norm} &= \frac{\sth}{1 + \sigma_{\rm det}^2 / \sigma_{\rm norm}^2} \cdot
           \frac{\sum_{A=N,F} \sum_{i=1}^n N_i^A \phase{i}{A}}
                {\sum_{A=N,F} \sum_{i=1}^n N_i^A}.
\end{align}
Thus, in the limit $N_i^A \rightarrow \infty$, $\chi^2$ is approximately given by
\begin{equation}
  \chi^2 = \sin^4 2 \theta_{13} \sum_{A=N,F} \sum_{i=1}^n N_i^A  \left(
              \frac{\sum_{B=N,F} \sum_{j=1}^n N_j^B \phase{j}{B}}
                   {\sum_{B=N,F} \sum_{j=1}^n N_j^B} - \phase{i}{A} \right)^2.
  \label{eq:chi2result}
\end{equation}
The $1\sigma$ sensitivity limit for $\sth$ is determined by the condition $\chi^2 = 1$:
\begin{equation}
  \sigma(\sth) = \left[ \sum_{A=N,F} \sum_{i=1}^n N_i^A \left(
              \frac{\sum_{B=N,F} \sum_{j=1}^n N_j^B [\phase{j}{B} - \phase{i}{A}]}
                   {\sum_{B=N,F} \sum_{j=1}^n N_{j}^{B}} \right)^2
      \right]^{-1/2} \ \stackrel{N_i^A \rightarrow \infty}{\longrightarrow} \ 0.
\end{equation}
This means that the sensitivity will be good if $N_i^A$ is large and 
$|\phase{j}{B} - \phase{i}{A}|$ is not too close to zero. Therefore an optimal
experiment involves one measurement very close to the reactor ($\phase{i}{A} \approx 0$)
and one around the first oscillation maximum ($\phase{i}{A} \approx 1$).

Now compare this to the case where the reactor position is fixed, i.e.\ $L^N = L^F$. 
Then, $N_i^N = N_i^F = N_i/2$ and \phase{i}{N} = \phase{i}{F} = \phase{i}{}. If 
we perform only a total rate analysis ($n = 1$), the right hand side of 
Eq.~\eqref{eq:chi2result} will be very close to zero. This means that we have to 
consider the next-order term:
\begin{equation}
  \chi^2 = \frac{a_{\rm norm}^2}{\sigma_{\rm norm}^2}
               + \frac{a_{\rm det}^2}{\sigma_{\rm det}^2}
         = \sin^4 2 \theta_{13} 
               \left( \frac{\sum_i N_i \phase{i}{}}{\sum_i N_i} \right)^2
             \frac{1}{\sigma_{\rm det}^2 + \sigma_{\rm norm}^2}.
\end{equation}
In this case, $\sigma(\sth)$ is of order $\sqrt{\sigma_{\rm det}^2 + \sigma_{\rm norm}^2}$.

In reality, $n$ will be chosen greater than~$1$ in order to exploit spectral 
information. In this case, the right hand side of Eq.~\eqref{eq:chi2result} can 
be interpreted as a comparison of the oscillation phase in bin $i$, \phase{i}{A}, 
to the average oscillation phase $\sum_B \sum_j N_j^B \phase{j}{B} / \sum_B \sum_j N_j^B$.
Therefore, this term is sensitive to the spectral distortion caused by neutrino
oscillations, and in principle, the systematical errors $\sigma_{\rm norm}$ and
$\sigma_{\rm det}$ can be eliminated without going to different baselines. This 
seems reasonable because the $L/E$ dependence of the oscillation probability 
Eq.~\eqref{eq:atmprob} implies that measuring at different energies is equivalent 
to measuring at different baselines. However, our numerical results presented 
below will show, that in reality it is still highly advantageous to change the 
baseline.

\subsection{Numerical results}

To demonstrate the full impact of systematical errors on the \sth\ sensitivity,
we have performed numerical simulations with a modified version of the {\sf GLoBES}
software \cite{globes}, which incorporates a $\chi^2$ expression similar to 
Eq.~\eqref{eq:chi2full}, but is based on the more realistic assumption of Poisson 
statistics and uses the exact three-flavor oscillation probability without 
applying any approximations. The reactor spectrum has been taken 
from~\cite{Murayama:2000iq,Eguchi:2002dm}, and the cross sections are 
from~\cite{Vogel:1999zy}.
We assume an energy resolution of $0.091 \sqrt{E / \textrm{MeV}}$~MeV
(\cite{Oberauer:2005kw, lothar}) and sort the events into 67 energy bins. 
Additionally, we assume that the reactor is located at baseline $L_1$ for 
$t_{\textrm{tot}} L_1^2 / (L_1^2 + L_2^2)$ and at baseline $L_2$ for 
$t_{\textrm{tot}} L_2^2 / (L_1^2 + L_2^2)$, where $t_{\textrm{tot}}$
is the total running time of the experiment. This ensures that the numbers of
events in the near and far detectors are comparable. We take $L_1 = 0.1$~km 
and $L_2 = 1.3$~km.

\begin{figure}
  \begin{center}
  \includegraphics[height=6 cm]{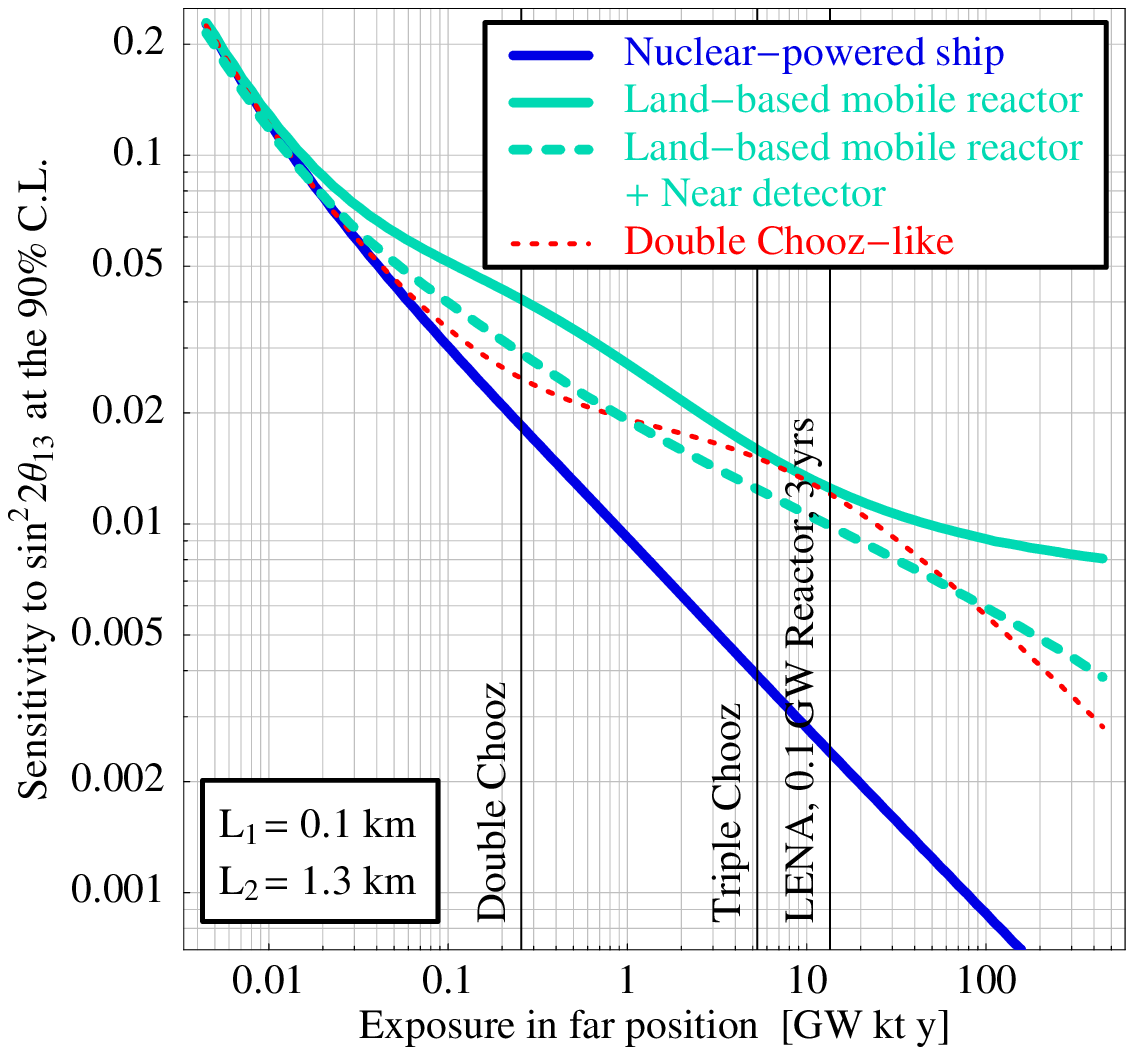}
  \includegraphics[height=6 cm]{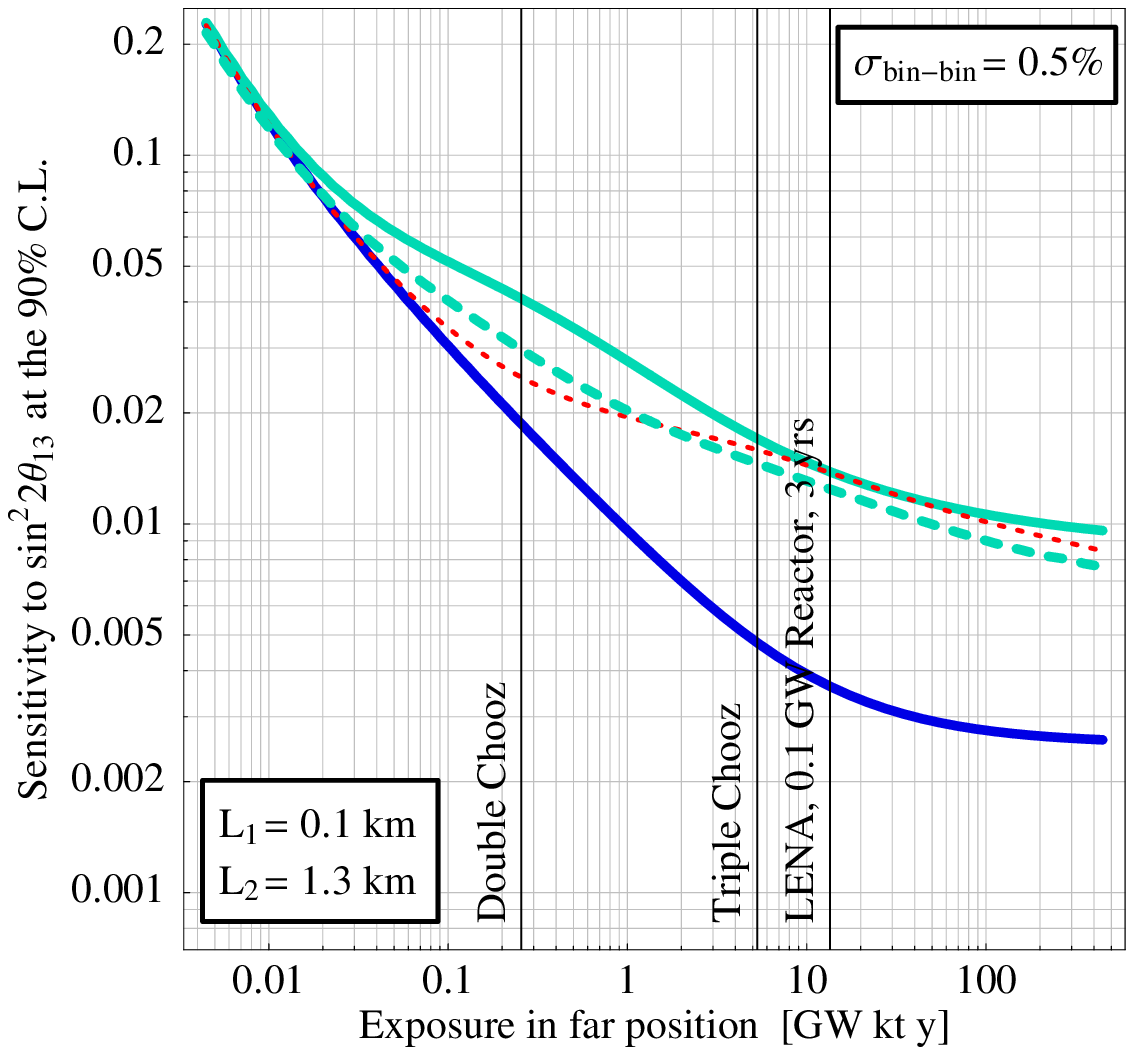}
  \end{center}
  \mycaption{\label{fig:exposure} Sensitivity to $\sin^22\theta_{13}$ at the 
  	90\% confidence level as a function of the exposure for a nuclear-powered 
	ship scenario (solid blue/black curve), a land-based mobile reactor scenario 
	(solid cyan/grey curve), a land-based mobile reactor scenario with an 
	associated near detector (dashed cyan/grey curve), and a Double Chooz-like 
	setup (dotted red/black curve) with one reactor and near and far detectors.
        The left plot has been calculated under the assumption of no bin-to-bin errors,
        while the right one includes an uncertainty $\sigma_{\rm bin-bin} = 0.5\%$.}
\end{figure}
Figure~\ref{fig:exposure} shows the sensitivity of different scenarios as a function
of the total exposure $\Phi$. In addition to the nuclear-powered ship and land-based 
scenarios that where discussed above, we also show for comparison the performance 
of a setup similar to Double Chooz~\cite{Ardellier:2004ui} with one reactor and 
two detectors. In this Double Chooz-like scenario, the uncorrelated detector 
normalization and energy calibration errors are taken to be only 0.6\% and 0.5\% 
because many detector-side effects will cancel. For very low exposures, 
$\Phi \leq 0.02$~GW~kt~y, all three scenarios are limited by the overall statistics 
and by backgrounds. As long as bin-to-bin errors, that are uncorrelated between the different bins 
as $\sigma_{\rm shape}$ and additionally uncorrelated between the near and far phase, are neglected (see left plot in
Fig.~\ref{fig:exposure}), the nuclear-powered ship scenario (solid blue/black curve)
follows the statistical limit also for larger exposures because systematical errors
cancel completely in this scenario, as expected from our analytical estimates.
For the land-based scenario (solid cyan/gray curve), only the systematical errors 
associated with the detector cancel because we have assumed the reactor neutrino 
flux to be completely uncorrelated before and after the displacement of the reactor. 
For the spectrum, we assume a partial correlation, which is parameterized by introducing 
in addition to the shape error from Eq.~\eqref{eq:chi2full} a new bin-dependent 
bias with a 1\% error, which is only present at baseline $L_2$. As soon as the 
exposure exceeds 0.02~GW~kt~y, the sensitivity of the land-based scenario becomes 
limited by the uncorrelated flux normalization error. As the exposure increases 
further, this error becomes less dominant because the event numbers are so large 
that spectral information can be exploited. However, above 10~GW~kt~y, the 
uncorrelated shape error prevents a further increase in the sensitivity. This 
can be avoided if a small near detector with a mass of 45~t, an uncorrelated
normalization error of 0.6\%, and an uncorrelated energy calibration error of 0.5\% 
is employed (dashed cyan/gray curve).
Finally, in the Double Chooz-like scenario (dotted red/black curve), the errors 
associated with the reactor cancel, but those associated with the detectors remain. 
As the reactor flux has larger uncertainties than the detector normalization, 
this scenario is better than the land-based mobile reactor scenario for 
luminosities of 0.02~GW~kt~y -- 6~GW~kt~y. At the onset of the spectrally dominated 
regime, the two curves meet again, but for very large luminosities the Double 
Chooz-like near/far scenario is again better than the land-based scenario (without 
associated near detector) because the shape error is canceled by the near detector.  

The vertical lines in Figure~\ref{fig:exposure} indicate the total exposures that are to
be expected from the Double Chooz experiment, its possible upgrade scenario Triple
Chooz (see \cite{Huber:2006vr}), and a possible total exposure of the mobile scenarios
(0.1~GW thermal power of the mobile reactor source, 3 years data taking, 45~kt fiducial
mass of the LENA detector). While the performance of the land-based mobile reactor
scenario is comparable to the Double Chooz-like scenario (dotted red/black curve) 
taken at the same total exposure, a nuclear-powered ship scenario could, due to the 
excellent cancellation of systematical uncertainties, achieve a sensitivity limit 
$\sin^22\theta_{13}\lesssim 0.003$ at the 90\% confidence level.

Even under the assumption of an uncorrelated bin-to-bin error 
$\sigma_{\rm bin-bin} = 0.5\%$ (see right hand plot of Fig~\ref{fig:exposure}),
this excellent sensitivity decreases only marginally. $\sigma_{\rm bin-bin}$
parameterizes all kinds of unknown backgrounds and detector non-linearities.
It is introduced in Eq.~\eqref{eq:chi2full} in a similar way as
$\sigma_{\rm shape}$, but the corresponding parameter $a_{{\rm bin-bin}, i}^A$
depends not only on the bin, but also on the detector position, hence the
index $A = N, F$ for near and far phase is introduced.

\begin{figure}
  \begin{center}
    \includegraphics[width=\textwidth]{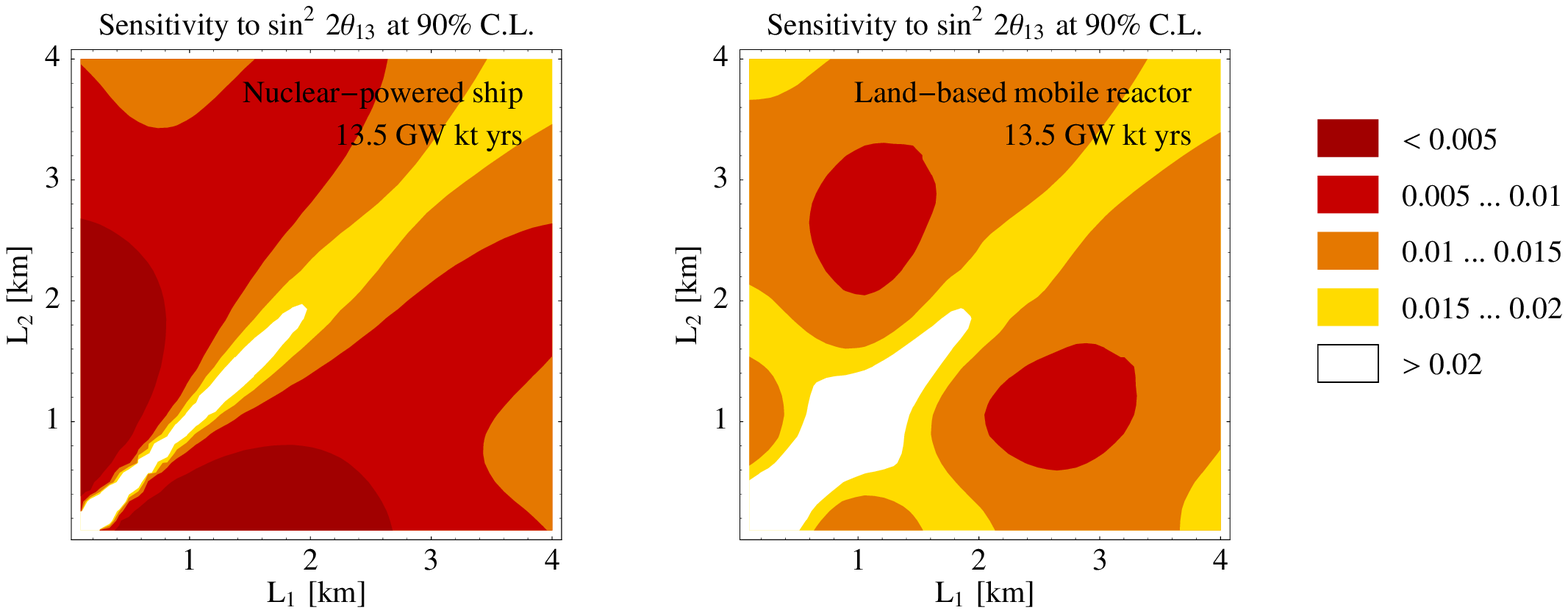}
  \end{center}
  \mycaption{\label{fig:bl} Sensitivity limit to $\sin^22\theta_{13}$ at the 90\% 
  	confidence level for the nuclear-powered ship scenario (left) and 
	the land-based scenario (right) as function of the near and far baselines.}
\end{figure}

Let us now address the issue of choosing the optimal baselines for the two phases
of the experiment. Figure~\ref{fig:bl} shows that, for the nuclear-powered ship
scenario, it is clearly advantageous to choose $L_1$ very close to the reactor 
in order to measure the unoscillated flux (0th oscillation minimum), and $L_2$ 
around the first oscillation maximum at about 1.3~km.
For the land-based scenario, the overall sensitivity is worse, but now the baseline 
combination 1st minimum/1st maximum gives better results than the combination 
0th minimum/1st maximum. From a practical point of view, this turns out to be 
advantageous because in the land-based scenario it will usually not be possible to 
move the reactor close to the detector, as the latter will be located deep underground.
We have also studied scenarios in which more than two different baselines are used. 
However, no significant improvement in the sensitivity could be achieved this way.

\section{Measurement of the solar parameters}
\label{sec:solar}

An often neglected topic is the issue of a precise measurement of the so-called 
solar oscillation parameters ($\theta_{12}$ and $\sdm$) with a reactor neutrino
experiment located at the first solar oscillation maximum (earlier studies can be found
in \cite{sado,Minakata:2005yb,Bandyopadhyay:2004cp}). Although e.g.\ the value 
of $\theta_{12}$ has been determined by solar experiments and KamLAND~\cite{Maltoni:2004ei} 
to be about $33^{\circ} \pm 3^{\circ}$, which seems to be quite accurate, it 
should be stressed that the relative uncertainty is still about 10\%. Since precise 
knowledge of $\theta_{12}$ is very important, in particular for the distinction 
between the normal and inverted neutrino mass hierarchy in 
$0\nu\beta\beta$-experiments~\cite{Choubey:2005rq,Lindner:2005kr} and via the 
day-night effect of solar neutrinos~\cite{Blennow:2003xw}, as well as for exploring 
the possible existence of quark-lepton complementarity~\cite{Minakata:2004xt}, a more 
precise measurement is desirable, and we will show that it can be performed with a
Large Liquid Scintillator Detector.

For the measurement of the solar oscillation parameters, two different scenarios 
are considered: the first one (``\emph{SMALL}'') employs a mobile reactor with a 
thermal power of 0.5~GW$_\mathrm{th}$, with 2 years of data taking, and the second 
one (``\emph{LARGE}'') is a power reactor (10 GW$_\mathrm{th}$, 5 years data 
taking) located at a suitable distance from the detector. It is desirable that 
this reactor should not be running during the whole lifetime of the detector, 
so that the latter can also pursue physics goals which require low backgrounds,
e.g.\ the search for Geo-neutrinos, supernova relic neutrinos, and proton decay. 
Of course, no one would build a detector like LENA voluntarily in the direct
neighborhood of such a power station, since the neutrino flux coming from this reactor would
dominate all events from other neutrino sources and the detector would be effectively
unusable for experiments other than long baseline oscillation measurements. However, if the
reactor is scheduled to be shut down after the first years of data taking with the Large Liquid Scintillator
Detector or if it is just planned to be built but there is 
enough time left to take data for the other measurements with the detector, this will not cause
any problems.

The detector properties and systematical errors are the same as in Section~\ref{sec:theta13},
but since background sources, in particular Geo-neutrinos, are much more important now,
they need to be treated in a more sophisticated way.

In Refs.~\cite{sado,Minakata:2005yb} it has been shown, that Geo-neutrinos have a
strong influence in a ``large reactor -- small detector setup'', namely SADO. Since we
analyze a much larger detector, this influence could be stronger here, even if
the ``product'' of the reactor and the detector size (i.e. the total exposure) is
similar. However, as will be shown, our results are still comparable to those
obtained by the previous analyses.

The $\chi^2$ function for Section 4, including contributions from Geo--neutrinos
in a more accurate way, has the following form:
\begin{multline}
  \chi^2 = \sum_{i} \frac{1}{N_i} \big[ T_i(a_{\rm norm}, a_{\rm det},
     a_{\rm reac}, a_{\rm U},a_{\rm Th}, b) - N_i \big]^2 +   \\
   + \frac{a_{\rm norm}^2}{\sigma_{\rm norm}^2} + \frac{a_{\rm det}^2}{\sigma_{\rm det}^2}
   + \frac{a_{\rm reac}^2}{\sigma_{\rm reac}^2}+ \frac{a_{\rm U}^2}{\sigma_{\rm U}^2}
   + \frac{a_{\rm Th}^2}{\sigma_{\rm Th}^2}
   + \frac{b^2}{\sigma_b^2}. \hspace*{2 cm}
   \label{eq:chi2full_solar}
\end{multline}
The meaning of the parameters is the same as in Sec.~\ref{sec:theta13}, but the
background errors $\sigma_{\rm bkg}$ are now split up into the contributions
coming from distant nuclear reactors ($\sigma_{\rm reac}$), the Geo-neutrinos
from the uranium decay chain ($\sigma_{\rm U}$), and the Geo-neutrinos from the
thorium decay chain ($\sigma_{\rm Th}$).

We will consider three different
situations to show the impact of the Geo-neutrino background:
\begin{itemize} 
  \item {\bf No Geo-neutrinos:} In this case, Geo-neutrinos are completely absent, and
    only the background from distant nuclear reactors is taken into account. As
    in Sec.~\ref{sec:theta13}, it yields 850 events per year. The uncertainty in the
    flux normalization is taken to be 2\%.
  \item {\bf Geo-neutrinos with a 10\% uncertainty:} Here, Geo-neutrinos are taken into
    account, and the uncertainty in their flux is assumed to be 10\%. We use independent
    normalization factors for neutrinos originating from the uranium decay chain and
    those originating from the thorium decay chain (cf.\ Eq.~\eqref{eq:chi2full_solar}) because the relative abundances of
    these elements in the Earth depend strongly on the geological
    model~\cite{Mantovani:2003yd}. However, as central value, we assume the ratio of the
    thorium and uranium abundances to be 3.9, as given in~\cite{Mantovani:2003yd}.   
    The decay chains of other long-lived radioactive isotopes, e.g.\ K-40, are not
    relevant for our discussion because they yield anti-neutrino energies below the 
    detection threshold of 1.8~MeV. The Geo-neutrino spectra in our simulations are taken
    from~\cite{GeoHP,Geothesis}. The reactor background is the same as in the scenario
    without Geo-neutrinos. 
  \item {\bf Geo-neutrinos with a 100\% uncertainty:} This scenario is equivalent to
    the previous one, but now the uncertainties in the two Geo-neutrino contributions
    are taken to be 100\%.
\end{itemize}

\subsection{Analytical estimates}

The optimum baseline for a measurement of $\theta_{12}$ can be estimated analytically 
if we neglect backgrounds and systematical errors for simplicity, and perform a
total rates analysis, neglecting spectral information. Again, we start with the
$\chi^2$-function for this scenario:
\begin {equation}
\chi^2=\frac{\left[ N \left(1-\sin^2 2\bar{\theta}_{12} \sin^2 \bar{\Delta} \right) - N
\left( 1-\sin^2 2\theta_{12} \sin^2 \Delta \right) \right]^2}{N \left( 1-\sin^2 2\theta_{12} \sin^2
\Delta \right)}.
  \label{eq:solarchi}
\end{equation}
Here, the barred values are the theoretical predictions, calculated with fit-values 
for the parameters, and the ones without bars come from the observed event rates, 
calculated with the assumed true parameter values from Eq.~\eqref{eq:oscparams}. 
The $\Delta$'s are defined as $\Delta \equiv \sdm L/(4E)$ and $\bar{\Delta} 
\equiv \overline{\Delta m^2_{21}} L/(4E)$, where $L$ denotes the baseline and $E$ 
the energy of the incoming neutrino. For convenience, we additionally introduce 
the abbreviations $s \equiv \sin 2\theta_{12}$ and $\bar{s} \equiv \sin 
2\bar{\theta}_{12}$, and write the unoscillated event rate as $N = N_0/L^2$, where
$N_0$ is independent of $L$. Furthermore, for the analytical estimate of the 
optimal baseline for a measurement of the solar mixing angle, we neglect 
parameter correlations between the solar parameters and assume a fixed 
$\sdm = \overline{\Delta m^2_{21}}$. Now, the $\chi^2$-function becomes:
\begin{equation}
\chi^2=\frac{N_0 \sin^4 \Delta}{L^2 \left( 1-s^2 \sin^2 \Delta \right)} \left(s^2-\bar{s}^2
\right)^2.
 \label{eq:theta12chi}
\end{equation}
Since we want to measure the amplitude $\sin^2 2\theta_{12}$ of the oscillations, 
we can already guess that the optimum baseline should correspond to an oscillation 
phase of approximately $\pi/2$, because there, the oscillation has a maximum, 
and $\pi/2$ is the closest point to the detector with this property. With a 
simple Taylor expansion, one then gets:
\begin{equation}
\sin \Delta =\sin \left( \frac{\pi}{2} + \left( \Delta -\frac{\pi}{2} \right) \right) \approx
1-\frac{1}{2} \left( \Delta - \frac{\pi}{2} \right)^2,
 \label{eq:solarTaylor}
\end{equation}
and hence, also up to leading order in the small quantity $\left( \Delta - \frac{\pi}{2}\right)$,
\begin{equation}
\sin^2 \Delta \approx 1-\left(\Delta - \frac{\pi}{2} \right)^2, \hspace{1.5 cm} \sin^4 \Delta \approx 1- 2
\left(\Delta - \frac{\pi}{2} \right)^2,
 \label{eq:furtherTaylor}
\end{equation}
which leads to
\begin{equation}
\chi^2 \approx \frac{N_0}{L^2} \frac{1-2 \left(\Delta - \frac{\pi}{2} \right)^2}{1-s^2 \left( 1-
\left(\Delta - \frac{\pi}{2} \right)^2 \right)} \left(\bar{s}^2-s^2\right)^2.
 \label{eq:approx12chi}
\end{equation}
Setting $\frac{\partial \chi^2}{\partial L}$ equal to zero, we get:
\begin{equation}
  -\frac{2}{L} \left[1-2\left(\Delta - \frac{\pi}{2} \right)^2 \right]
   - \frac{\sdm}{E} (\Delta - \frac{\pi}{2}) - \frac{\sdm}{2E} (\Delta -\frac{\pi}{2})s^2
\frac{1-2\left(\Delta-\frac{\pi}{2}\right)^2}{1-s^2 \left(1-\left(\Delta -\frac{\pi}{2} \right)^2
\right)}=0.
 \label{eq:chi12zero}
\end{equation}
To lowest order in $(\Delta-\frac{\pi}{2})$, this leads to a baseline of
\begin{equation}
L_{\textrm{OPT}} \approx \frac{\pi E}{\sdm}\left( 1 \pm \sqrt{1-\frac{8}{\pi^2 (1+\frac{1}{2}t^2)}}
\right),
 \label{eq:best12bl}
\end{equation}
where $t\equiv \tan 2\theta_{12}$. Only the plus-solution is self-consistent with 
our initial assumption $\Delta \approx \pi/2$. For $E = 4$~MeV, it yields a best
baseline of approximately 55 km.

For the determination of $\sdm$, the optimum baseline will be somewhat different, 
but in this case, the calculation is not as simple because $L$ appears both inside 
and outside the $\sin\Delta$ terms, so $\sdm-\overline{\Delta m^2_{21}}$ cannot 
be extracted. Approximations are quite cumbersome because of the complex interplay 
of the different terms, and spectral effects complicate the calculation even 
further. One can however estimate that the best baseline should correspond to an 
oscillation phase $\Delta$ for which small variations of $\sdm$ will have the 
strongest effect on the oscillation probability. It can be read off from 
Eq.~\eqref{eq:solprob} that this will be the case if $\Delta$ is an odd integer 
multiple of $\pi/4$: $\Delta = \pi/4, 3\pi/4, 5\pi/4, \dots$. Low multiples are
favored by higher statistics (due to the geometrical factor $L^{-2}$), while for 
higher multiples the effect is larger because a variation of $\sdm$ implies a 
stretching of the $\sin^2$ function in Eq.~\eqref{eq:solprob}, which is 
proportional to $L$. This discussion shows that reliable estimates for the 
optimal baseline can only be obtained numerically.

\begin{figure}
  \begin{center}
    \includegraphics[width=13cm]{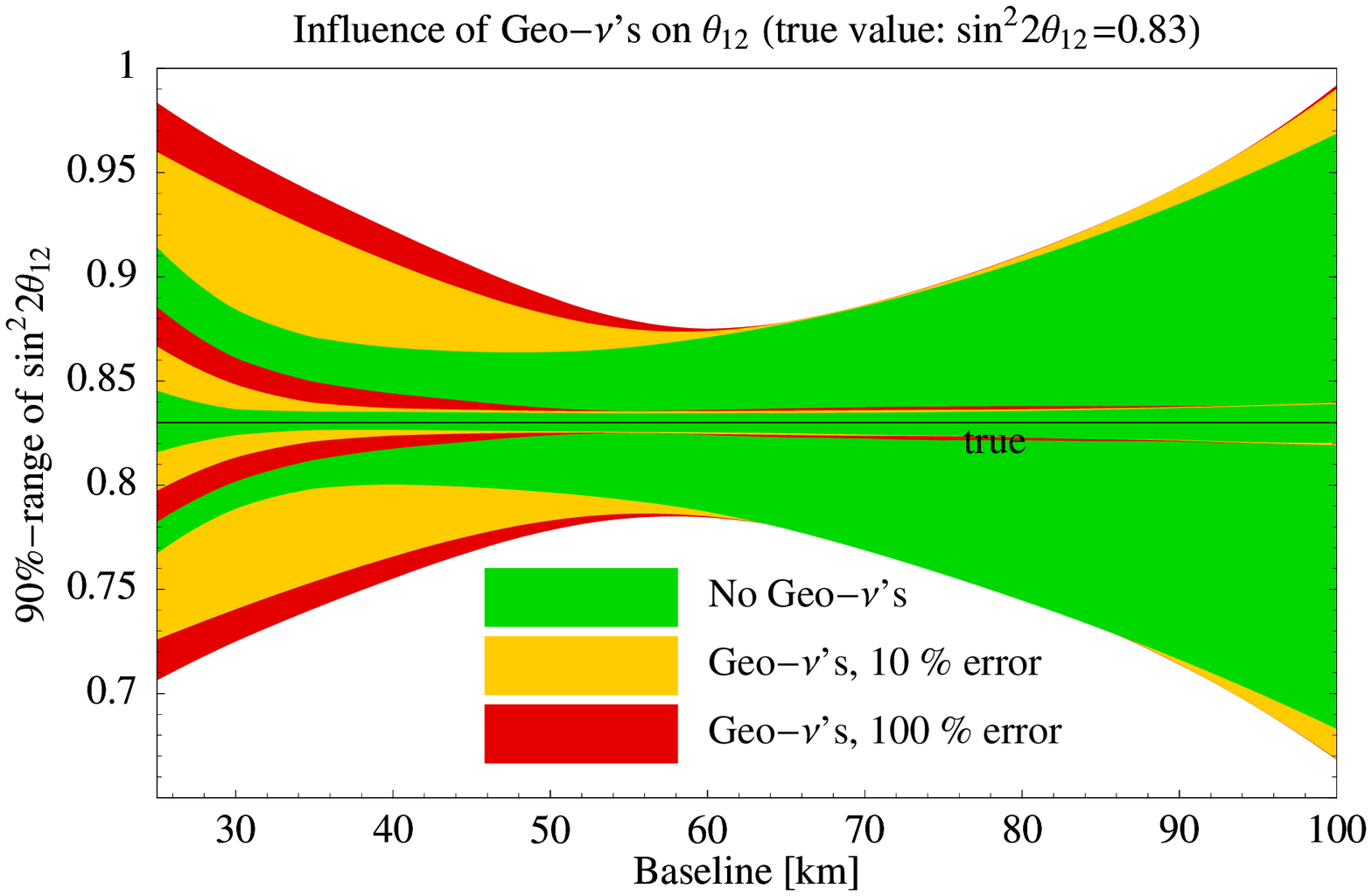} \\
    
    \vspace*{0.5cm}
    \includegraphics[width=13cm]{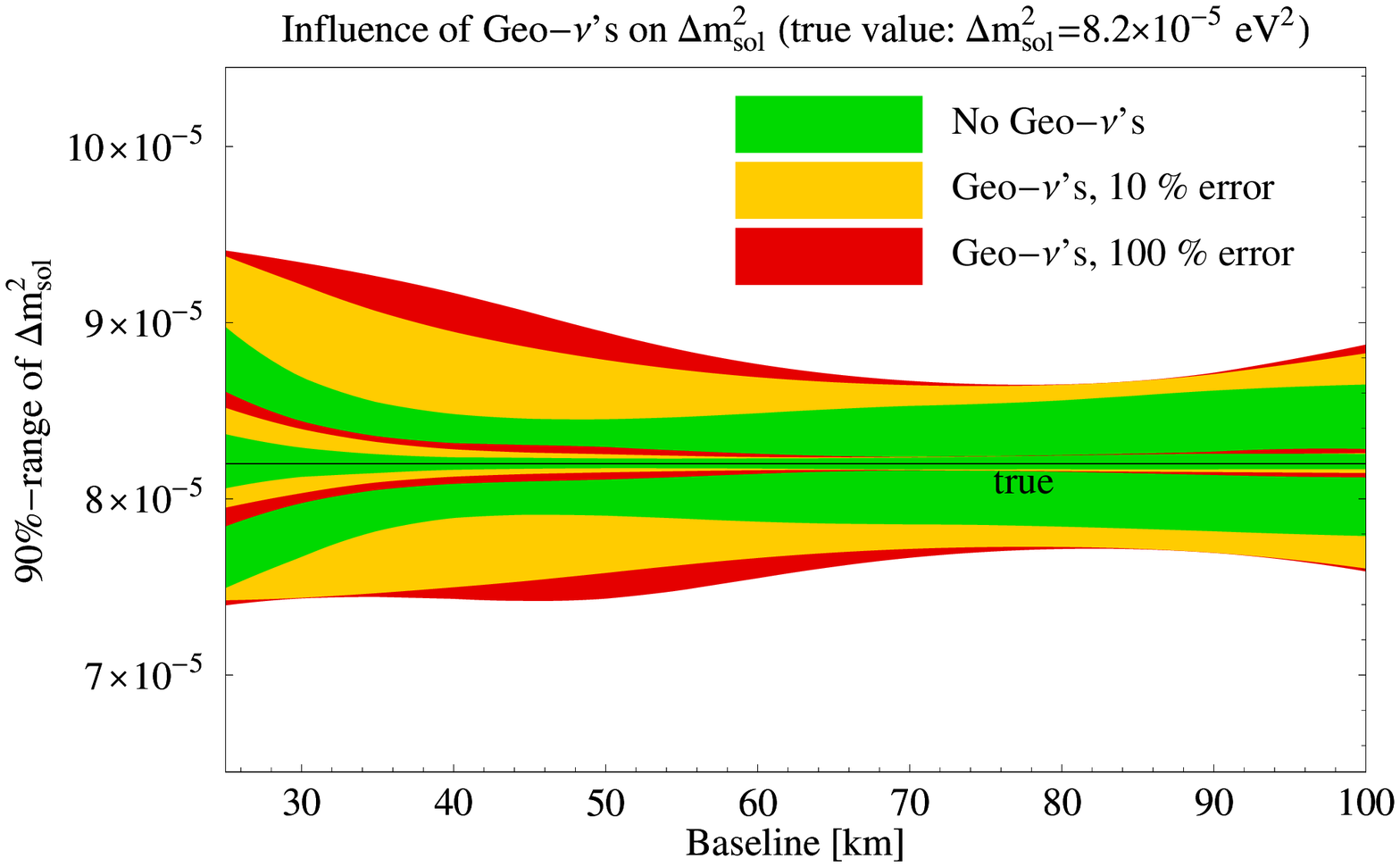}
  \end{center}
  \mycaption{\label{fig:baseline} Achievable precision for the parameters 
  	$\sin^22\theta_{12}$ (upper panel) and $\sdm$ (lower panel) at 
	the 90\% confidence level as a function of the baseline. The outer
 	curves correspond to the \emph{SMALL} scenario, while the inner 
	ones show the achievable precision withe the \emph{LARGE} scenario. 
	The true values we have chosen are indicated by the horizontal lines.}
\end{figure}

\subsection{Numerical results}

Looking at Figure~\ref{fig:baseline}, one can see that the optimum baselines for a
measurement of the solar mixing angle $\sin^22\theta_{12}$ for the \emph{SMALL} 
scenario will be around 50 to 70~km if one includes Geo-neutrinos, and a bit 
shorter without Geo-neutrinos (approximately 40--60~km). This is in agreement 
with the optimum baselines obtained for the SADO scenario in
Refs.~\cite{sado,Minakata:2005yb} (where the technically different situation
``large reactor -- small detector'' is investigated) and 
with our analytical estimates. Geo-neutrinos only perturb the low energy part of 
the spectrum up to around 3.3~MeV, which means that, if they are present, the 
overall sensitivity will be dominated by the high energy part. However, for high 
energy neutrinos the optimum baseline will be larger due to the $L/E$-dependence 
of the oscillation probabilities. For the \emph{LARGE} scenario, all baselines 
are more or less equivalent, as long as one is far enough away from the reactor 
to be able to resolve the solar oscillation,~i.e.\ the oscillatory disappearance 
``dip'' in the spectrum has to be seen. 
For the determination of $\sdm$, Figure~\ref{fig:baseline} and Table~\ref{tab:solar} 
show that the optimum baseline turns out to be 47.0~km for the \emph{SMALL} and 
53.2~km for the \emph{LARGE} scenario, as long as the Geo-neutrino background 
is neglected. With the inclusion of this background, the best baseline for the 
\emph{SMALL} scenario is shifted to 75--80~km, while for the \emph{LARGE} scenario, 
the impact of the background is less dramatic.

\begin{table}
  \begin{center}
  \begin{tabular}{llrrr} \hline
$\sin^2 2\theta_{12}$ &  & $\mathrm{L_{OPT}\, [km]}$ & events/year & precision (90\% CL) \\ \hline
 \emph{SMALL} & no-Geo $\nu$'s & 43.5 & 1866 (1026) & +4.10\%/-3.61\% \\ 
  & 10\% error & 57.5 & 2853 (514) & +5.30\%/-5.30\%  \\ 
  & 100\% error & 59.4 & 2828 (488) & +5.42\%/-5.42\%  \\ \hline
 \emph{LARGE}  & no-Geo $\nu$'s & 40.0 & 27572 (26732) & +0.60\%/-0.36\%  \\
  & 10\% error & 52.9 & 14405 (12065) & +0.60\%/-0.60\% \\ 
  & 100\% error & 55.0 & 13466 (11127) & +0.72\%/-0.60\%  \\ \hline 
  \end{tabular}

\vspace*{0.2cm}
  \begin{tabular}{lcccr} \hline
$\sdm$ &  & $\mathrm{L_{OPT}\, [km]}$ & events/year & precision  (90\% CL) \\ \hline
 \emph{SMALL} & no Geo-$\nu$'s & 47.0 & 1655 (815) & +3.05\%/-3.54\%  \\ 
  & 10\% error & 76.2 & 2729 (389) & +5.37\%/-5.73\%   \\ 
  & 100\% error & 79.7 & 2717 (378) & +5.49\%/-5.85\% \\ \hline
 \emph{LARGE}  & no Geo-$\nu$'s & 53.2 & 12757 (11917) & +0.37\%/-0.37\%  \\
  & 10\% error & 65.0 & 11116 (8776) & +0.37\%/-0.37\%   \\ 
  & 100\% error & 70.0 & 10589 (8250) & +0.49\%/-0.49\%  \\ \hline 
  \end{tabular}
  \end{center}
  \mycaption{\label{tab:solar} The achievable precision on the solar parameters 
  	$\sin^2 2\theta_{12}$ and $\sdm$ at the 90\% confidence level. Also shown 
	are the optimal baselines and total (signal) event rates per year for the 
	measurements in the \emph{LARGE} and \emph{SMALL} scenarios. Note that for 
	the \emph{LARGE} scenario, the exact baseline is not so important since 
	one always has extremely high rates.}
\end{table}
To get a better overview and to get a clue of how strongly the different scenarios 
are influenced by Geo-neutrinos, we summarize the numerical results in Table~\ref{tab:solar}. 
Taking into account, that our values give the 90\% confidence level ranges for 
the solar parameters, one can easily see that already the \emph{SMALL} scenario could 
reach values comparable to SADO~\cite{sado,Minakata:2005yb}. It is important that, for a shorter 
baseline, the impact of Geo-neutrinos broadens the 90\% confidence level range 
quite strongly, even if the flux error is only taken to be 10\%. Hence, for 
performing a reactor experiment at such baselines, it is necessary to have a good 
knowledge of the Geo-neutrino fluxes. This goal can be reached for example by 
measuring the Geo-neutrino background before turning on or after turning off the 
mobile reactor. However, for the \emph{LARGE} scenario, one can clearly see that 
an amazing precision in the measurement of the solar parameters is possible, even 
without precise knowledge on the Geo-neutrino background. One could of course assume, that such a precision could be spoiled by the fact that the energy of a
single neutrino cannot be determined better than about $1$ or $2\%$. However, note that
this precision is just the limit for each single bin. Since we consider a total of 67 bins
the overall precision can indeed be better. More precisely,
$2\%$ is the \emph{absolute} uncertainty for each bin, but since information from different bins 
is compared, the \emph{relative} uncertainty between the bins is the main limiting factor and this can, due
to the spectral information, be much less than the former one. Hence, the precision of our
results is realistic for the considered scenarios.

\section{Conclusions}

We have performed analytical and numerical calculations to estimate the potential 
of reactor anti-neutrino disappearance measurements with a Large Liquid Scintillator 
Detector like LENA. For the measurement of the small mixing angle $\sin^22\theta_{13}$, 
we have assumed the reactor to be mobile. This allows subsequent measurements in 
a near and a far position, so that many systematical uncertainties are canceled. 
We have distinguished between two different scenarios, a nuclear-powered ship 
(e.g.\ an icebreaker or a submarine) and a land-based mobile reactor, where the 
cancellation of systematical uncertainties is different. In the case of a
nuclear-powered ship, detector related and source related systematical errors are 
eliminated, so that the $\sin^22\theta_{13}$ sensitivity follows the statistical 
limit with increasing exposure. In the case of the land-based mobile reactor, 
the source related flux normalization error remains because the reactor needs to 
be shut down for the displacement. If an exposure of 13.5~GW~kt~yrs is assumed, 
the nuclear-powered ship scenario can provide a limit of $\sin^22\theta_{13} 
\lesssim 0.004$ at the 90\% confidence level, whereas the land-based scenario can 
achieve a limit of $\sin^22\theta_{13} \lesssim 0.02$ with the same exposure. 
Furthermore, we have shown that for the nuclear-powered ship scenario the two
baselines, near and far, should be chosen around the 0th oscillation minimum
($L \approx 0$~km) and 1st oscillation maximum ($L \approx 1.3$~km), while the
land-based scenario yields optimal results with near and far baselines around the
1st oscillation minimum and 1st oscillation maximum. This is advantageous because 
the detector will be located underground, and the mobile reactor cannot be positioned 
in the direct neighborhood of the detector. 

We have also analyzed the potential of a Large Liquid Scintillator Detector to 
perform precision measurements of the solar parameters $\sin^22\theta_{12}$ and 
$\Delta m^2_{21}$. For this, we have compared two scenarios: a small 0.5~GW reactor
(\emph{SMALL}) and a large 10~GW power station (\emph{LARGE}). As has previously been shown
in~\cite{sado}, the measurement of the solar parameters is strongly
influenced by the Geo-neutrino background, therefore we have implemented an accurate
treatment of this background, and compared its impact for different assumptions on
the systematical uncertainties. We have shown, that the \emph{SMALL} scenario would
favor baselines of approximately 40~to~60~km for the measurement of $\sin^22\theta_{12}$
and 47~km for the measurement of $\Delta m^2_{21}$. The exact choice of the baseline
in the \emph{LARGE} scenario turned out to be less crucial, since the excellent
statistics provide precise information on the energy spectrum. Even with the most
conservative assumptions on the systematical uncertainties of the Geo-neutrino
background, the \emph{SMALL} scenario can already achieve precisions of approximately
5.4\% on $\sin^22\theta_{12}$ and 5.9\% on $\Delta m^2_{21}$, while the
\emph{LARGE} scenario achieves precisions of approximately 0.7\% on
$\sin^22\theta_{12}$ and 0.5\% on $\Delta m^2_{21}$. This very good accuracy is
required e.g.\ to test quark-lepton complementarity at a very high precision level.

\section*{Acknowledgments}

We would like to thank K.~Hochmuth, P.~Huber, T.~Marrod\'an-Undagoitia,
L.~Oberauer, and M.~Wurm for useful discussions and information on the LENA 
detector. We are also grateful to S.~Enomoto for providing Geo-neutrino 
spectra in machine-readable form. This work has been supported by the 
``Sonderforschungsbereich 375 f\"ur Astro-Teilchenphysik der Deutschen 
Forschungsgemeinschaft'' and the Graduiertenkolleg 1054. JK would like to 
acknowledge support from the Studienstiftung des Deutschen Volkes.

\bibliographystyle{./apsrev}
\bibliography{./mobile-reactor}

\end{document}